\documentclass[prd,aps,nofootinbib,eqsecnum]{revtex4}
\usepackage{graphicx,color,amsmath,amsxtra}
\usepackage{amssymb}
\usepackage{enumerate}
\usepackage{hhline}
\usepackage{array}
\usepackage{tabularx}

\newcommand{\bear}{\begin{array}}  \newcommand{\eear}{\end{array}}
\newcommand{\bea}{\begin{eqnarray}}  \newcommand{\eea}{\end{eqnarray}}
\newcommand{\beq}{\begin{equation}}  \newcommand{\eeq}{\end{equation}}
\newcommand{\bef}{\begin{figure}}  \newcommand{\eef}{\end{figure}}
\newcommand{\bec}{\begin{center}}  \newcommand{\eec}{\end{center}}

\newcommand{\Eqn}[1]{&\hspace{-0.2em}#1\hspace{-0.2em}&}

\def\Vec#1{\mbox{\boldmath $#1$}}

\def\be{\begin{equation}}
\def\ee{\end{equation}}
\def\bea{\begin{eqnarray}}
\def\eea{\end{eqnarray}}
\def\beq{\begin{eqnarray}}
\def\eeq{\end{eqnarray}}
\def\tr{{\rm tr}\, }
\def\nn{\nonumber \\}
\def\e{{\rm e}}

\begin{document}

\title{
Cosmology in non-minimal Yang-Mills/Maxwell theory
}

\author{
Kazuharu Bamba$^{1}$ and Shin'ichi Nojiri$^2$}
\affiliation{
$^1$Department of Physics, National Tsing Hua University, Hsinchu, Taiwan 300\\
$^2$Department of Physics, Nagoya University, Nagoya 464-8602, Japan
}


\begin{abstract}
We review cosmology in non-minimal Yang-Mills/Maxwell theory, 
in which the Yang-Mills/electromagnetic field couples to a function of the 
scalar curvature. We show that power-law inflation can be realized due to 
the non-minimal gravitational coupling of the Yang-Mills field which may be 
caused by quantum corrections. Moreover, we study non-minimal vector model 
in the framework of modified gravity and demonstrate that both inflation and 
the late-time accelerated expansion of the universe can be realized. 
We also discuss the cosmological reconstruction of the Yang-Mills theory. 
Furthermore, we investigate late-time cosmology in non-minimal 
Maxwell-Einstein theory. We explore the forms of the non-minimal gravitational 
coupling which generate the finite-time future singularities and 
the general conditions for this coupling in order that the finite-time 
future singularities cannot appear. 
\end{abstract}


\maketitle

\section{Introduction}

It is observationally confirmed that the current expansion of the universe is 
accelerating, as well as that there existed the inflationary stage 
in the early universe~\cite{WMAP1, SN1}. The former phenomenon is 
called ``dark energy'' problem 
(for recent reviews, see~\cite{DE-rev, Nojiri:2006ri}). 
The studies of the dark energy problem can be categorized into the following 
two approaches. 
One is to introduce some (unknown) matter which is responsible for 
dark energy in the framework of general relativity. 
Another is to modify the gravitational theory, e.g., to study the action 
described by an arbitrary function $F(R)$ of the scalar curvature $R$, 
which is called ``$F(R)$ gravity'' (for a review, see~\cite{Nojiri:2006ri, 
MG-rev}). 
Such a modified gravity must pass cosmological bounds and solar system tests
because it is considered as an alternative gravitational theory.

A very realistic modified gravitational theory which evades 
solar-system tests has recently been proposed
by Hu and Sawicki~\cite{Hu:2007nk}
(for related studies, see~\cite{related studies}).
In this theory, an effective epoch described by the cold dark matter model
with cosmological constant ($\Lambda$CDM), which accounts for high-precision
observational data, is realized as in general relativity with 
cosmological constant. 
Although this theory can successfully explain the late-time acceleration
of the universe, the possibility of the realization of inflation has not 
been discussed. 
In Refs.~\cite{Nojiri:2007cq, viable modified gravities, Nojiri:2007as}, 
following the previous inflation-acceleration 
unification proposal~\cite{Nojiri:2003ft}, 
modified gravities in which both inflation and the late-time acceleration 
of the universe can occur have been investigated. The classification of viable 
$F(R)$ gravities has also been suggested 
in Ref.~\cite{viable modified gravities}. 

As another gravitational source of inflation and the 
late-time acceleration of the universe, there is a coupling between the scalar 
curvature and matter Lagrangian~\cite{matter-1, Allemandi:2005qs} 
(see also~\cite{Deruelle:2008fs}). 
Such a coupling may be applied for the realization of the dynamical 
cancellation of cosmological constant~\cite{DC}. 
In Refs.~\cite{criteria-1, Faraoni:2007sn, Bertolami:2007vu}, 
the criteria for the viability of such theories have been examined. 
As a simple case, a coupling between a function of 
the scalar curvature and the kinetic term of a massless scalar field 
in a viable modified gravity has been considered in Ref.~\cite{Nojiri:2007bt}. 

It is known that the coupling between the scalar curvature and the Lagrangian 
of the electromagnetic field arises in curved spacetime due to one-loop 
vacuum-polarization effects in 
Quantum Electrodynamics (QED)~\cite{Drummond:1979pp}. 
In the present article, following the considerations 
in Ref.~\cite{Bamba:2008xa}, 
we review cosmology in non-minimal non-Abelian gauge theory 
(Yang-Mills (YM) theory), in which the non-Abelian gauge field 
(the YM field) couples to a function of the scalar curvature. 
In particular, it is shown that power-law inflation can be realized due to 
the non-minimal gravitational coupling of the Yang-Mills field. 
The consequences presented correspond to the generalization of the results for 
non-minimal Maxwell theory with the coupling of the electromagnetic field 
to a function of the scalar curvature~\cite{Bamba:2008ja}. 
Furthermore, we consider a non-minimal vector model in the framework of 
modified gravity. It is demonstrated that both inflation and 
the late-time accelerated expansion of the universe can be realized. 
We also study the reconstruction of the YM theory. 
In the past studies, inflation driven by a vector 
filed~\cite{vector inflation} and 
its instability~\cite{Himmetoglu:2008zp} and 
gravitational-electromagnetic inflation from a 5-dimensional vacuum 
state~\cite{G-M inflation from 5D} have been discussed. 
As a candidate for dark energy, 
the effective YM condensate~\cite{YM-DE}, 
the Born-Infeld quantum condensate~\cite{Elizalde:2003ku} 
and a vector field~\cite{vector field DE} have been proposed. 
Models of vector curvaton have also been constructed~\cite{vector curvaton}. 

In addition, following the investigations in Ref.~\cite{Bamba:2008ut}, 
we review late-time cosmology in the non-minimal Maxwell-Einstein theory. 
We investigate the forms of the non-minimal gravitational coupling 
of the electromagnetic field generating the finite-time future singularities 
and the general conditions for the non-minimal gravitational coupling of 
the electromagnetic field in order that the finite-time future singularities 
cannot appear. In Ref.~\cite{Hollenstein:2008hp}, 
$F(R)$ gravity coupled to non-linear electrodynamics has been examined. 

This article is organized as follows. 
In Sec.~II we investigate a non-minimal gravitational coupling of 
the $SU(N)$ YM field in general relativity. 
In Sec.~III we consider non-minimal vector model in the framework 
of modified gravity. 
In Sec.~IV we discuss the reconstruction of the YM theory. 
In Sec.~V we study non-minimal Maxwell-Einstein theory with 
a general gravitational coupling. We explore the cosmological effects of 
the non-minimal gravitational coupling of the electromagnetic field to a 
function of the scalar curvature on the finite-time future singularities. 
Finally, summary is given in Sec.~VI.
We use units in which $k_\mathrm{B} = c = \hbar = 1$ and denote the
gravitational constant $8 \pi G$ by ${\kappa}^2$, so that
${\kappa}^2 \equiv 8\pi/{M_{\mathrm{Pl}}}^2$, where
$M_{\mathrm{Pl}} = G^{-1/2} = 1.2 \times 10^{19}$GeV is the Planck mass.
Moreover, in terms of electromagnetism we adopt Heaviside-Lorentz units.

\section{Inflation in general relativity}

In this section, we study non-minimal YM theory in general relativity. 
The model action is given as follows~\cite{Bamba:2008xa}:
\begin{eqnarray}
S_{\mathrm{GR}} \Eqn{=}
\int d^{4}x \sqrt{-g}
\left( 
{\mathcal{L}}_{\mathrm{EH}}
+{\mathcal{L}}_{\mathrm{YM}} \right)\,,
\label{eq:2.1} \\
{\mathcal{L}}_{\mathrm{EH}}
\Eqn{=}
\frac{1}{2\kappa^2} R\,,
\label{eq:2.2} \\
{\mathcal{L}}_{\mathrm{YM}}
\Eqn{=}
-\frac{1}{4} I(R) F_{\mu\nu}^{a}F^{a\mu\nu}
\left[1+b\tilde{g}^2 \ln
\left| \frac{-\left( 1/2 \right) F_{\mu\nu}^{a}F^{a\mu\nu}}{\mu^4} \right|
\right]\,,
\label{eq:2.3} \\
I(R) \Eqn{=} 1+f(R)\,, \quad 
b = \frac{1}{4} \frac{1}{8 \pi^2}\frac{11}{3}N\,, \quad 
F_{\mu\nu}^{a} =
{\partial}_{\mu}A_{\nu}^{a} - {\partial}_{\nu}A_{\mu}^{a}
+ f^{abc} A_{\mu}^{b} A_{\nu}^{c}\,,
\label{eq:2.6}
\end{eqnarray}
where $g$ is the determinant of the metric tensor $g_{\mu\nu}$,
$R$ is the scalar curvature arising from the spacetime
metric tensor $g_{\mu\nu}$,
and ${\mathcal{L}}_{\mathrm{EH}}$ is the Einstein-Hilbert action.
Moreover, ${\mathcal{L}}_{\mathrm{YM}}$ with $I(R)=1$ is the effective
Lagrangian of the $SU(N)$ YM theory up to one-loop
order~\cite{YM-L, Adler:1983zh},
$f(R)$ is an arbitrary function of $R$,
$b$ is the asymptotic freedom constant,
$F_{\mu\nu}^{a}$ is the field strength tensor,
$A_{\mu}^{a}$ is the $SU(N)$ YM field
with the internal symmetry index $a$
(Roman indices, $a$, $b$, $c$, run over $1, 2, \ldots, N^2-1$, and
in $F_{\mu\nu}^{a}F^{a\mu\nu}$ the summation in terms of the index $a$ is
also made), and
$f^{abc}$ is a set of numbers called structure constants and completely
antisymmetric~\cite{P-S}.
Furthermore, $\mu$ is the mass scale of the renormalization point, and
a field-strength-dependent running coupling constant
is given by~\cite{Adler:1983zh}
\begin{eqnarray}
\tilde{g}^2 (X) 
= \frac{\tilde{g}^2}{1+b\tilde{g}^2\ln \left| X/ \mu^4 \right|}\,, 
\quad 
X \equiv -\frac{1}{2} F_{\mu\nu}^{a} F^{a\mu\nu}\,, 
\label{eq:2.8}
\end{eqnarray}
where $\tilde{g}$ is the value of the running coupling constant
when $X = \mu^4$.

Taking the variations of the action~(\ref{eq:2.1}) with respect to the metric 
$g_{\mu\nu}$ and the $SU(N)$ YM field $A_{\mu}^{a}$, we obtain the 
gravitational field equation and the equation of motion of $A_{\mu}^{a}$ as 
\begin{eqnarray}
\hspace{-5mm}
R_{\mu \nu} - \frac{1}{2}g_{\mu \nu}R
\Eqn{=} \kappa^2 T^{(\mathrm{YM})}_{\mu \nu}\,,
\label{eq:2.9} \\
T^{(\mathrm{YM})}_{\mu \nu}
\Eqn{=}
I(R) \left( g^{\alpha\beta} F_{\mu\beta}^{a} F_{\nu\alpha}^{a} \varepsilon
-\frac{1}{4} g_{\mu\nu} \mathcal{F} \right)
+\frac{1}{2} \left[ f^{\prime}(R) \mathcal{F} R_{\mu \nu}
+ g_{\mu \nu} \Box \left( f^{\prime}(R) \mathcal{F} \right)
- {\nabla}_{\mu} {\nabla}_{\nu}
\left( f^{\prime}(R) \mathcal{F} \right)
\right]\,,
\label{eq:2.10} \\
\varepsilon \Eqn{=}
1+b\tilde{g}^2 \ln \left| e
\left[ \frac{-\left( 1/2 \right) F_{\mu\nu}^{a}F^{a\mu\nu}}{\mu^4}
\right] \right|
= 1+b\tilde{g}^2 \ln \left| e \left( \frac{X}{\mu^4} \right)
\right|\,,
\label{eq:2.11} \\
\mathcal{F} \Eqn{=}
F_{\mu\nu}^{a}F^{a\mu\nu}
\left[ 1+b\tilde{g}^2 \ln \left|
\frac{-\left( 1/2 \right) F_{\mu\nu}^{a}F^{a\mu\nu}}{\mu^4} \right|
\right]
= -2X\left( 1+b\tilde{g}^2 \ln \left| \frac{X}{\mu^4} \right| \right)\,,
\label{eq:2.12} \\
&&
\hspace{-34.5mm}
\mathrm{and} 
\hspace{15mm}
\frac{1}{\sqrt{-g}}{\partial}_{\mu}
\left( \sqrt{-g} I(R) \varepsilon
F^{a\mu\nu} \right)
-I(R) \varepsilon f^{abc} A_{\mu}^{b} F^{c\mu\nu}
= 0\,,
\label{eq:2.13}
\end{eqnarray}
respectively, where $T^{(\mathrm{YM})}_{\mu \nu}$ is the contribution to 
the energy-momentum tensor from the $SU(N)$ YM field, 
the prime denotes differentiation with respect to $R$, 
${\nabla}_{\mu}$ is the covariant derivative operator associated with
$g_{\mu \nu}$, 
$\Box \equiv g^{\mu \nu} {\nabla}_{\mu} {\nabla}_{\nu}$
is the covariant d'Alembertian for a scalar field, 
and $R_{\mu \nu}$ is the Ricci curvature tensor. 
Moreover, $\varepsilon$ is a field-strength-dependent effective dielectric 
constant~\cite{Adler:1983zh}, and $e \approx 2.72$ is the Napierian number. 
In deriving the second equalities in 
Eqs.~(\ref{eq:2.11}) and (\ref{eq:2.12}), we have used 
$X = -\left(1/2\right) F_{\mu\nu}^{a} F^{a\mu\nu}$. 

We assume the flat Friedmann-Robertson-Walker (FRW) spacetime 
with the metric 
$
{ds}^2 =-{dt}^2 + a^2(t)d{\Vec{x}}^2
= a^2(\eta) ( -{d \eta}^2 + d{\Vec{x}}^2 )
$, 
where $a$ is the scale factor and $\eta$ is the conformal time.

In the search of exact solutions for non-minimal YM (electromagnetic)-gravity 
theory, the problem of off-diagonal components of YM (electromagnetic) stress 
tensor being non-zero while the right-hand side of Einstein equations is zero 
(for the argument about the problem of off-diagonal components of 
electromagnetic energy-momentum tensor in non-minimal Maxwell-gravity 
theory, see~\cite{OD}). 
As a simple case, we can consider the following case in which 
the off-diagonal components of $T^{(\mathrm{YM})}_{\mu \nu}$ 
in Eq.~(\ref{eq:2.10}) vanishes: 
(i) Only (YM) magnetic fields exist and hence (YM) electric fields are 
negligible. 
(ii) $\Vec{B^{a}} = (B^{a}_1, B^{a}_2, B^{a}_3)$, where 
$B^{a}_1 = B^{a}_2 =0, B^{a}_3 \neq 0$, namely,
we consider the case in which only one component of $\Vec{B}^{a}$ is non-zero 
and other two components are zero. 
In such a case, it follows from $\mathrm{div} \Vec{B}^{a} = 0$ that 
the off-diagonal components of the last term 
on the right-hand side of 
$T^{(\mathrm{YM})}_{\mu \nu}$, i.e., 
${\nabla}_{\mu} {\nabla}_{\nu} \left( f^{\prime}(R) \mathcal{F} \right)$
are zero. 
Thus, all of the off-diagonal components of $T^{(\mathrm{YM})}_{\mu \nu}$ 
are zero. 
Throughout this article, we consider the above case. 

In Eq.~(\ref{eq:2.13}), 
we can neglect the higher order than or equal to the quadratic terms in 
$A_{\mu}^{a}$ because the amplitude of $A_{\mu}^{a}$ is small, 
and investigate the linearized equation of 
Eq.\ (\ref{eq:2.13}) in terms of $A_{\mu}^{a}$ in the Coulomb gauge, 
${\partial}^jA_j^{a}(t,\Vec{x}) =0$, and the case of 
$A_{0}^{a}(t,\Vec{x}) = 0$. 
Replacing the independent variable $t$ by $\eta$, we find that
the Fourier mode $A_i^{a}(k,\eta)$ satisfies the equation 
$
\left(\partial^2 A_i^{a}(k,\eta)\right)/\left(\partial \eta^2\right) + 
\left(1/I(\eta)\right) \left[d I(\eta)/\left(d \eta\right) \right] 
\left[ \partial A_i^{a}(k,\eta)/\left(\partial \eta\right) \right]
+ k^2 A_i^{a}(k,\eta) = 0
$. 
By using the WKB approximation on subhorizon scales and the long-wavelength 
approximation on superhorizon scales, and matching these solutions at the 
horizon crossing~\cite{Bamba-mag-WKB}, we find that an approximate 
solution is given by 
\begin{eqnarray}
\hspace{-5mm}
\left|A_i^{a}(k,\eta)\right|^2
= |C(k)|^2
= \frac{1}{2kI(\eta_k)}
\left|1- \left( \frac{1}{2kI(\eta_k)}\frac{d I(\eta_k)}{d \eta}
+ i \right)k\int_{\eta_k}^{{\eta}_{\mathrm{f}}}
\frac{I(\eta_k)}{I \left(\tilde{\eta} \right)}
d\tilde{\eta}\,\right|^2\,,
\label{eq:A-4}
\end{eqnarray}
where $\eta_k$ and ${\eta}_{\mathrm{f}}$ are the conformal time 
at the horizon-crossing and at the end of inflation, respectively. 
It follows from 
$
{B_i}^{\mathrm{proper}}(t,\Vec{x}) = 
a^{-2}{\epsilon}_{ijk}{\partial}_j A_k(t,\Vec{x})
$, where ${\epsilon}_{ijk}$ is the totally antisymmetric tensor
(${\epsilon}_{123}=1$), that the amplitude of 
the proper YM magnetic fields on a comoving scale $L=2\pi/k$ 
in the position space is given by 
\begin{eqnarray}
|{B_i^{a}}^{(\mathrm{proper})}(t)|^2 = 
\frac{k|C(k)|^2}{\pi^2}\frac{k^4}{a^4}
\left[
1 + \frac{1}{2} f^{abc} u^{b} u^{c} \frac{k|C(k)|^2}{2 \pi^2}
\right]\,,
\label{eq:A-5}
\end{eqnarray} 
where $u^{b}(=1)$ and $u^{c}(=1)$ are the quantities denoting the dependence 
on the indices $b$ and $c$, respectively. From Eq.~(\ref{eq:A-5}), we see that 
the YM magnetic fields evolves as 
$|{B_i^{a}}^{(\mathrm{proper})}(t)|^2 = |\bar{B}^{a}|^2/a^4$, where 
$|\bar{B}^{a}|$ is a constant. 

In this case, 
using the $(\mu,\nu)=(0,0)$ component and 
the trace part of the $(\mu,\nu)=(i,j)$ component of Eq.~(\ref{eq:2.9}),
where $i$ and $j$ run from $1$ to $3$, and 
eliminating $I(R)$ from these equations, we obtain
\begin{eqnarray}
\hspace{-10mm}
\dot{H} + \frac{\varepsilon}{\varepsilon - b\tilde{g}^2} H^2 
\Eqn{=} \kappa^2
\biggl(
f^{\prime}(R) \left\{ -(2 \varepsilon + b\tilde{g}^2)\dot{H}
+ \left[ \left( \frac{3\varepsilon - 7b\tilde{g}^2}
{\varepsilon - b\tilde{g}^2} \right) \varepsilon + 8b\tilde{g}^2
\right] H^2
\right\}
+3f^{\prime\prime}(R)
\Bigl[ (\varepsilon - b\tilde{g}^2)\dddot{H}
\nonumber \\
&&
\hspace{-2mm}
{}- 2\left(\varepsilon + 2b\tilde{g}^2\right)H\ddot{H} 
+4(\varepsilon - b\tilde{g}^2)\dot{H}^2 - 24 \varepsilon H^2\dot{H}
\Bigr]
+18f^{\prime\prime\prime}(R) (\varepsilon - b\tilde{g}^2)
\left( \ddot{H} + 4H\dot{H} \right)^2
\biggr) 
\frac{|\bar{B}^{a}|^2}{a^4}\,,
\label{eq:2.48}
\end{eqnarray}
where $H=\dot a/a$ is the Hubble parameter and a dot denotes a time 
derivative, $\dot{~}=\partial/\partial t$. From Eq.~(\ref{eq:2.11}), 
we see that the value of $\varepsilon$ 
depends on the field strength, in other words, it varies in time.
The change in time of $\varepsilon$, however, is smaller than 
that of other quantities 
because the dependence of $\varepsilon$ on the field strength is 
logarithmic, so that we can approximately regard $\varepsilon$ as 
constant in Eq.~(\ref{eq:2.48}). 
In what follows, we regard $\varepsilon$ as constant.

We consider the case in which $f(R)$ is given by the 
Hu-Sawicki form~\cite{Hu:2007nk, viable modified gravities}, 
\begin{eqnarray}
f(R) = f_{\mathrm{HS}}(R) \equiv \frac{c_1 \left(R/m^2 \right)^n}
{c_2 \left(R/m^2 \right)^n + 1}\,,
\label{eq:2.49}
\end{eqnarray}
which satisfies the conditions: 
$
\lim_{R\to\infty} f_{\mathrm{HS}}(R) = c_1/c_2 = \mbox{const} 
$ 
and 
$
\lim_{R\to 0} f_{\mathrm{HS}}(R) = 0. 
$ 
Here, $c_1$ and $c_2$ are dimensionless constants, $n$ is a positive 
constant, and $m$ denotes a mass scale. 
The above second condition means that there could exist a flat 
spacetime solution. 
Hence, because in the late time universe the value of the scalar curvature
becomes zero, the YM coupling $I$ becomes unity, so
that the standard YM theory can be naturally recovered. 
To show that power-law inflation can be realized,
we consider the case in which the scale factor is given by
$a(t) = \bar{a} \left(t/\bar{t}\right)^p$,
where $\bar{t}$ is some fiducial time during inflation,
$\bar{a}$ is the value of $a(t)$ at $t=\bar{t}$,
and $p$ is a positive constant.
In this case, $H=p/t$ and 
$R=6\left(\dot{H}+2H^2\right)=6p(2p-1)/t^2$. 
At the inflationary stage, because $R/m^2 \gg 1$, we can use the approximate 
relation 
$
f_{\mathrm{HS}}(R) \approx 
\left(c_1/c_2\right) \left[1-\left(1/c_2\right) 
\left( R/m^2 \right)^{-n} \right].
$ 
Substituting the above relations in terms of $a$, $H$ and $R$, and 
the approximate expressions of 
$f_{\mathrm{HS}}^{\prime}(R)$, 
$f_{\mathrm{HS}}^{\prime\prime}(R)$ and
$f_{\mathrm{HS}}^{\prime\prime\prime}(R)$ derived from 
the above approximate expression of $f_{\mathrm{HS}}(R)$ for $R/m^2 \gg 1$ 
into Eq.~(\ref{eq:2.48}), we find 
$p = \left(n+1\right)/2$. 
If $n \gg 1$, $p$ becomes much larger than unity, so that 
power-law inflation can be realized. Consequently, 
the YM field with a non-minimal gravitational 
coupling in Eq.~(\ref{eq:2.3}) can be a source of inflation. 

The constraint on a non-minimal gravitational coupling of matter from
the observational data of the central temperature of the Sun has been
proposed~\cite{Bertolami:2007vu}.
Furthermore, the existence of the non-minimal gravitational coupling of
the electromagnetic field changes the
value of the fine structure constant, i.e., the strength of the
electromagnetic coupling. Hence, the deviation of the non-minimal
electromagnetism from the ordinary Maxwell theory can be constrained
from the observations of radio and optical quasar absorption
lines~\cite{Tzanavaris:2006uf}, those of the anisotropy of the cosmic
microwave background (CMB) radiation~\cite{Battye:2000ds, Stefanescu:2007aa},
those of the absorption of the CMB radiation at 21 cm hyperfine transition of 
the neutral atomic hydrogen~\cite{Khatri:2007yv}, 
and big bang nucleosynthesis (BBN)~\cite{Bergstrom:1999wm, Avelino:2001nr}
as well as solar-system experiments~\cite{Fujii:2006ic}
(for a recent review, see~\cite{GarciaBerro:2007ir}).
On the other hand, because the energy scale of the YM theory is higher than
the electroweak scale, the existence of the non-minimal gravitational coupling 
of the YM field might influence on models of the grand unified theories (GUT).

\section{Non-minimal vector model}

In this section, we investigate cosmology in non-abelian non-minimal 
vector model in the framework of $F(R)$ gravity. 
The model action is given as follows~\cite{Bamba:2008xa}:
\begin{eqnarray}
\bar{S}_{\mathrm{MG}}
\Eqn{=}
\int d^{4}x \sqrt{-g}
\left( 
{\mathcal{L}}_{\mathrm{MG}}
+{\mathcal{L}}_{\mathrm{V}} \right)\,,
\label{eq:4.1} \\
{\mathcal{L}}_{\mathrm{MG}}
\Eqn{=}
\frac{1}{2\kappa^2} \left( R+F(R) \right)\,,
\label{eq:3.2} \\
{\mathcal{L}}_{\mathrm{V}}
\Eqn{=}
I(R) \left(
-\frac{1}{4} F_{\mu\nu}^{a} F^{a\mu\nu} - V\left[A^{a2}\right]
\right)\,,
\label{eq:4.2}
\end{eqnarray} 
where $F(R)$ is an arbitrary function of $R$, 
$F_{\mu\nu}^{a}$ is given by Eq.~(\ref{eq:2.6}), and
$A^{a2}= g^{\mu \nu} A_{\mu}^{a} A_{\nu}^{a}$. 
($F(R)$ is the modified part of gravity, and hence 
$F(R)$ is completely different from the non-minimal gravitational 
coupling of the YM field $f(R)$ in (\ref{eq:2.6}).)

We should note that the last term $V\left[A^{a2}\right]$ 
in the action (\ref{eq:4.2}) is 
not gauge invariant but can be rewritten in a gauge invariant way. 
For example, if the gauge group is a unitary group, we may introduce a 
$\sigma$-model like field $U$, 
which satisfies $U^\dagger U=1$. The last term could be rewritten in the 
gauge invariant form: 
\begin{eqnarray}
\label{sn1}
V\left[A^{a2}\right] \to 
V\left[ -\bar{c} \tr \left(U^\dagger D_{\mu} U\right)
\left(U^\dagger D^{\mu} U\right) \right]\,,
\end{eqnarray}
where $\bar{c}$ is a constant for the normalization and $D_\mu$ is a 
covariant derivative $D_\mu = \partial_\mu + iA_\mu^a T^a$ ($T^a$'s are 
the generators of the gauge algebra). 
If we choose the unitary gauge $U=1$, the term in (\ref{sn1}) reduces to 
the original one: $V\left[A^{a2}\right]$. This may tells that the action 
(\ref{eq:4.2}) described 
the theory where the gauge group is spontaneously broken.

Taking the variations of the action~(\ref{eq:4.1}) with respect to the metric 
$g_{\mu\nu}$ and the vector field $A_{\mu}^{a}$, we obtain the 
gravitational field equation and the equation of motion of $A_{\mu}^{a}$ as 
\begin{eqnarray}
&&
\hspace{-15mm}
\left( 1+F^{\prime}(R) \right) R_{\mu \nu}
- \frac{1}{2}g_{\mu \nu} \left( R+F(R) \right) + g_{\mu \nu}
\Box F^{\prime}(R) - {\nabla}_{\mu} {\nabla}_{\nu} F^{\prime}(R)
= \kappa^2 T^{(\mathrm{V})}_{\mu \nu}\,,
\label{eq:4.3} \\
T^{(\mathrm{V})}_{\mu \nu}
\Eqn{=}
I(R) \left( g^{\alpha\beta} F_{\mu\beta}^{a} F_{\nu\alpha}^{a}
+ 2 A_{\mu}^{a} A_{\nu}^{a}
\frac{d V\left[A^{a2}\right]}{d A^{a2}}
-\frac{1}{4} g_{\mu\nu} \bar{\mathcal{F}} \right) 
\nonumber \\
&&
{}+\frac{1}{2} \left\{ f^{\prime}(R) \bar{\mathcal{F}} R_{\mu \nu}
+ g_{\mu \nu} \Box \left[ f^{\prime}(R) \bar{\mathcal{F}} \right]
- {\nabla}_{\mu} {\nabla}_{\nu}
\left[ f^{\prime}(R) \bar{\mathcal{F}} \right]
\right\}\,,
\label{eq:4.4} \\
\bar{\mathcal{F}} 
\Eqn{=} 
F_{\mu\nu}^{a} F^{a\mu\nu} + 4V\left[A^{a2}\right]\,,
\label{eq:4.5} \\
&&
\hspace{-45mm}
\mathrm{and} 
\hspace{25mm}
\frac{1}{\sqrt{-g}}{\partial}_{\mu}
\left( \sqrt{-g} I(R) F^{a\mu\nu} \right)
- I(R) \left(
f^{abc} A_{\mu}^{b} F^{c\mu\nu}
+ 2 \frac{d V\left[A^{a2}\right]}{d A^{a2}} A^{a \nu}
\right)
= 0\,,
\label{eq:4.6}
\end{eqnarray}
respectively, where $T^{(\mathrm{V})}_{\mu \nu}$ is the contribution to 
the energy-momentum tensor from $A_{\mu}^{a}$.

We consider the case in which $V\left[A^{a2}\right]$ 
is given by a class of the following power-law potential: 
$
V\left[A^{a2}\right]=\bar{V} \left( A^{a2}/\bar{m}^2 
\right)^{\bar{n}}
$, 
where $\bar{V}$ is a constant, $\bar{m}$ denotes a mass scale, and 
$\bar{n} (>1)$ is a positive integer. 
Similarly to the preceding section, we consider the linearized equation 
of Eq.~(\ref{eq:4.6}) in terms of $A_{\mu}^{a}$. 
For the above power-law potential, 
the form of the linearized equation of motion under the ansatz 
${\partial}^jA_j^{a}(t,\Vec{x}) =0$ and $A_{0}^{a}(t,\Vec{x}) = 0$ 
is the same as in the preceding section. 
(This is similar to the Coulomb gauge but since the action (\ref{eq:4.2}) is 
not gauge invariant, or gauge symmetry is completely fixed by the unitary 
gauge as in after (\ref{sn1}), this condition is only a working hypothesis.) 

  From $A_{0}^{a} = 0$, we have 
$\left( 1/a^2 \right) A_{i}^{a} A_{i}^{a} d V\left[A^{a2}\right]/ 
\left( d A^{a2} \right)
= \bar{n} V\left[A^{a2}\right]$. 
Using the $(\mu,\nu)=(0,0)$ component and 
the trace part of the $(\mu,\nu)=(i,j)$ component of Eq.~(\ref{eq:4.3}), 
and eliminating $I(R)$ from these 
equations, we obtain 
\begin{eqnarray}
&& \hspace{-8mm}
\dot{H} + H^2 +
\left\{ \frac{1}{6} F(R) - F^{\prime}(R) H^2 +
3F^{\prime\prime}(R) \left[
\dddot{H} + 4\left( \dot{H}^2 + H\ddot{H} \right) \right]
+ 18F^{\prime\prime\prime}(R) \left( \ddot{H} + 4H\dot{H} \right)^2
\right\} 
\nonumber \\
&& 
{}=
\kappa^2 \biggl(
\biggl[
f^{\prime}(R) \left( -2\dot{H} + 3H^2 \right) +
3f^{\prime\prime}(R) \left( \dddot{H}-2H\ddot{H}+4\dot{H}^2-24H^2\dot{H}
\right) 
+18f^{\prime\prime\prime}(R)
\left( \ddot{H} + 4H\dot{H} \right)^2
\biggr]
\frac{|\bar{B}^{a}|^2}{a^4} 
\nonumber \\
&& \hspace{10mm}
{}
+2 \biggl\{
-f^{\prime}(R) \left[ \bar{n} \dot{H} +
\left( 1 + 2 \bar{n} - 2\bar{n}^2 \right) H^2
\right]
+3f^{\prime\prime}(R) \left[
\dddot{H} + 2\left( 3-2\bar{n} \right) H\ddot{H} + 4\dot{H}^2
+ 8\left( 1-2\bar{n} \right) H^2\dot{H}
\right] 
\nonumber \\
&& \hspace{10mm}
{}+18f^{\prime\prime\prime}(R) \left( \ddot{H} + 4H\dot{H} \right)^2
\biggr\} V\left[A^{a2}\right]
\biggr)\,.
\label{eq:4.13}
\end{eqnarray} 
In the case in which 
$|{B_i^{a}}^{(\mathrm{proper})}(t)|^2 = |\bar{B}^{a}|^2/a^4$, 
$V\left[A^{a2}\right] \propto a^{-2\bar{n}}$. If $\bar{n}=2$, 
the time evolution of $V\left[A^{a2}\right]$ is the same as that of 
$|{B_i^{a}}^{(\mathrm{proper})}(t)|^2$. On the other hand, if
$\bar{n} \geq 2$, $V\left[A^{a2}\right]$ decreases much more rapidly than 
$|{B_i^{a}}^{(\mathrm{proper})}(t)|^2$ during inflation. 
Hence, in the latter case we can neglect the terms proportional to 
$V\left[A^{a2}\right]$ on the right-hand side of Eq.~(\ref{eq:4.13}). 

We examine the following case. 
$F(R)$ is given by~\cite{Nojiri:2007as}
\begin{eqnarray}
F(R) \Eqn{=}
- M^2 \frac{\left[ \left(R/M^2\right) - \left(R_0/M^2\right) \right]^{2l+1}
+ {\left(R_0/M^2\right)}^{2l+1}}
{c_3 + c_4 \left\{
\left[ \left(R/M^2\right) - \left(R_0/M^2\right) \right]^{2l+1} +
{\left(R_0/M^2\right)}^{2l+1} \right\}}\,,
\label{eq:3.5}
\end{eqnarray}
which satisfies the conditions:
$
\lim_{R\to\infty} F(R) = -M^2/c_4 = \mbox{const}
$, 
$
\lim_{R\to 0} F(R) = 0.
$
Here, $c_3$ and $c_4$ are dimensionless constants,
$l$ is a positive integer, and $M$ denotes a mass scale.
We consider that in the limit $R\to\infty$, i.e., at the very early stage of 
the universe, $F(R)$ becomes an effective cosmological constant, 
$
\lim_{R\to\infty} F(R) = -M^2/c_4 = -2{\Lambda}_{\mathrm{i}}, 
$ 
where 
${\Lambda}_{\mathrm{i}} \left(\gg {H_0}^2 \right)$ is an effective
cosmological constant in the very early universe, 
and that at the present time $F(R)$ becomes a small constant, 
$
F(R_0) = -M^2 \left(R_0/M^2\right)^{2l+1}/ 
\left[ {c_3 + c_4 \left(R_0/M^2\right)^{2l+1}} \right] = -2R_0, 
$
where $R_0 \left(\approx {H_0}^2 \right)$ is current curvature. 
Here, 
$H_0$ is the Hubble constant at the present time:
$H_{0} = 100 h \hspace{1mm} \mathrm{km} \hspace{1mm} {\mathrm{s}}^{-1}
\hspace{1mm} {\mathrm{Mpc}}^{-1}
= 2.1 h \times 10^{-42} {\mathrm{GeV}}
\approx 1.5 \times 10^{-33} {\mathrm{eV}}$~\cite{Kolb and Turner}, where
we have used $h=0.70$~\cite{Freedman:2000cf}. 
Moreover, $f(R)$ is given by~\cite{Nojiri:2007as}:
\begin{eqnarray}
f(R) = f_{\mathrm{NO}}(R) \equiv
\frac{\left[ \left(R/M^2\right) - \left(R_0/M^2\right) \right]^{2q+1}
+ {\left(R_0/M^2\right)}^{2q+1}}
{c_5 + c_6 \left\{
\left[ \left(R/M^2\right) - \left(R_0/M^2\right) \right]^{2q+1} +
{\left(R_0/M^2\right)}^{2q+1} \right\}}\,,
\label{eq:3.10}
\end{eqnarray}
which satisfies the following conditions:
$
\lim_{R\to\infty} f_{\mathrm{NO}}(R) = 1/c_6 = \mbox{const},
$
$
\lim_{R\to 0} f_{\mathrm{NO}}(R) = 0.
$
Here, $c_5$ and $c_6$ are dimensionless constants, and
$q$ is a positive integer. 
The form of $F(R)$ in Eq.~(\ref{eq:3.5}) and that of 
$f_{\mathrm{NO}}(R)$ in Eq.~(\ref{eq:3.10}) 
correspond to the extension of that of $f_{\mathrm{HS}}(R)$ in 
Eq.~(\ref{eq:2.49}). It has been shown in Ref.~\cite{Nojiri:2007as} 
that modified gravitational theories described by 
the action (\ref{eq:3.2}) with $F(R)$ in Eq.~(\ref{eq:3.5}) successfully 
pass the solar-system tests as well as cosmological bounds, 
and that they are free of instabilities. 

Using Eqs.~(\ref{eq:4.13}), (\ref{eq:3.5}) and (\ref{eq:3.10}), 
we find that at the very early stage of the universe 
($R/M^2 \gg 1$ and $R/M^2 \gg R_0/M^2$), 
$
a(t) \propto \exp \left(\sqrt{{\Lambda}_{\mathrm{i}}/3} t \right), 
$ 
so that exponential inflation can be realized, 
and that at the present time ($F(R) = F(R_0) = -2R_0$), 
$ 
a(t) \propto \exp \left(\sqrt{R_0/3} t \right), 
$ 
so that the late-time acceleration of the universe can also be realized.

\section{Reconstruction of the YM theory}

In this section, we indicate how to reconstruct the YM theory from the 
known evolution of the universe (for a review, see~\cite{Nojiri:2006be}). 
We consider the following action: 
\be
\label{YM1}
S=\int d^4 x \sqrt{-g}\left(\frac{R}{2\kappa^2} + \tilde{\cal F}
\left(F^a_{\mu\nu}F^{a\,\mu\nu}\right)
\right)\ .
\ee
By introducing an auxiliary scalar field $\phi$, we may rewrite the action
(\ref{YM1}) in the
following form:
\be
\label{YM2}
S=\int d^4 x \sqrt{-g}\left(\frac{R}{2\kappa^2} + \frac{1}{4}P(\phi)
F^a_{\mu\nu}F^{a\,\mu\nu}
+ \frac{1}{4}Q(\phi) \right)\ .
\ee
Taking the variations of the action~(\ref{YM2}) with respect to $\phi$, 
we obtain 
\be
\label{YM3}
0= \frac{d P(\phi)}{d \phi} F^a_{\mu\nu}F^{a\,\mu\nu} + 
\frac{d Q(\phi)}{d \phi}\ ,
\ee
which could be solved with respect $\phi$ as
$\phi=\phi\left(F^a_{\mu\nu}F^{a\,\mu\nu}\right)$. 
Here, the prime denotes differentiation with respect to $\phi$. 
Substituting the expression into the action (\ref{YM2}), we obtain the action 
(\ref{YM1}) with 
\be
\label{YM4}
\tilde{\cal F}\left(F^a_{\mu\nu}F^{a\,\mu\nu}\right) = \frac{1}{4}\left(
P\left(\phi\left(F^a_{\mu\nu}F^{a\,\mu\nu}\right)\right)
F^a_{\mu\nu}F^{a\,\mu\nu}
+ Q\left(\phi\left(F^a_{\mu\nu}F^{a\,\mu\nu}\right)\right) \right)\ .
\ee
Taking the variations of the action (\ref{YM2}) with respect to the 
metric tensor $g_{\mu\nu}$, we obtain the Einstein equation:
\be
\label{YM5}
\frac{1}{2\kappa^2}\left( R_{\mu\nu} - \frac{1}{2}R g_{\mu\nu} \right)
= - \frac{1}{2} P(\phi) F^a_{\mu\rho} F^{a\ \rho}_{\ \nu}
+ \frac{1}{8}g_{\mu\nu} \left( P(\phi) F^a_{\rho\sigma}F^{a\,\rho\sigma} +
Q(\phi) \right)\ .
\ee
On the other hand, taking the variations of the action (\ref{YM2}) with 
respect to $A^a_\mu$, we obtain 
\be
\label{YM6}
0=\partial_\nu \left( \sqrt{-g}P(\phi)F^{a\,\nu\mu}\right)
 - \sqrt{-g} P(\phi) f^{abc} A^b_\nu F^{c\,\nu\mu}\ .
\ee
For simplicity, we only consider the case in which the gauge algebra is 
$SU(2)$, where $f^{abc}=\epsilon^{abc}$, and we assume that the gauge 
fields are given in the following form: 
\be
\label{YM7}
A^a_\mu = \left\{ \begin{array}{cl}
\bar{\alpha}\e^{\lambda(t)}\delta^a_{\ i} & (\mu=i=1,2,3) \\
0 & (\mu=0)
\end{array} \right. \ , 
\ee
where $\bar{\alpha}$ is a constant with mass dimension and
$\lambda$ is a proper function of $t$.
In general, if the vector field is condensed, the rotational invariance
of the universe could be broken.
In case of (\ref{YM7}), the direction of the vector field is gauge variant. 
Hence, all the gauge invariant quantities given by (\ref{YM7}) do not break 
the rotational invariance.

By the assumption, Eq.~(\ref{YM3}) has the following form:
\be
\label{YM8}
0= 6 \left( - \bar{\alpha}^2 {\dot \lambda}^2 \e^{2\lambda} a^{-2} +
\bar{\alpha}^4 \e^{4\lambda} a^{-4} \right)
\frac{d P(\phi)}{d \phi} + \frac{d Q(\phi)}{d \phi}\ ,
\ee
and $(t,t)$-component of Eq.~(\ref{YM5}) is given by
\be
\label{YM9}
0=\frac{3}{\kappa^2}H^2 - \frac{3}{2}\left( \bar{\alpha}^2
{\dot \lambda}^2 \e^{2\lambda} a^{-2}
+ \bar{\alpha}^4 \e^{4\lambda} a^{-4} \right) - \frac{1}{4}Q(\phi)\ .
\ee
The $\mu=0$ component of Eq.~(\ref{YM6}) becomes identity and $\mu=i$ 
component gives
\be
\label{YM10}
0= \partial_t \left(a P(\phi) \dot \lambda \e^{\lambda} \right)
 - 2 \bar{\alpha}^2 a^{-1} P(\phi) \e^{3\lambda}\ .
\ee

Since we can always take the scalar field $\phi$ properly, we may identify the 
scalar field with the time coordinate $\phi=t$. 
By differentiating Eq.~(\ref{YM9}) with respect to $t$ and eliminating
$\dot Q= d Q(\phi)/\left( d \phi \right)$, we obtain
\bea
\label{YM11}
0 = \frac{2}{\kappa^2}H\dot H + \bar{\alpha}^2 {\dot\lambda}^2 \e^{2\lambda}
a^{-2} \dot P - P\left[ \bar{\alpha}^2
\left(\dot\lambda \ddot\lambda + {\dot\lambda}^3 - \lambda^2 H\right)
\e^{2\lambda} a^{-2} + 2 \bar{\alpha}^4 \left( \dot\lambda - H \right)
\e^{4\lambda} a^{-4} \right]\ .
\eea
Furthermore, eliminating $\dot P$ by using Eq.~(\ref{YM10}), we find
\be
\label{YM12}
P=\frac{2H\dot H}{\kappa^2 \left[ 2\bar{\alpha}^2 a^{-2} \e^{2\lambda}
\left({\dot\lambda}^2
+ \dot\lambda \ddot\lambda \right) - \bar{\alpha}^4 \e^{4\lambda} a^{-4} H
\right] }\ .
\ee
Using Eq.~(\ref{YM12}), we can eliminate $P$ (and $\dot P$) in 
Eq.~(\ref{YM10}) and obtain
\bea
\label{YM13}
0 \Eqn{=} 2 \left(\dot\lambda \dddot\lambda + {\ddot\lambda}^2
+ 3{\dot\lambda}^2 \ddot\lambda \right) - \bar{\alpha}^2
\e^{2\lambda} a^{-2} \dot H
+ 4\left({\dot\lambda}^3 + \dot\lambda \ddot\lambda
 - \bar{\alpha}^2
\e^{2\lambda} a^{-2} H \right)\left(\dot\lambda - H\right) \nn
&& + \left[2 \left({\dot\lambda}^3 + \dot\lambda \ddot\lambda \right)
- \bar{\alpha}^2 \e^{2\lambda} a^{-2} H \right] \left(\frac{\dot H}{H} +
\frac{\ddot H}{\dot H}
+ H + \frac{\ddot\lambda}{\dot\lambda} + \dot\lambda
 - \frac{2\bar{\alpha}^2 a^{-2}\e^{2\lambda}}{\dot\lambda} \right)\ .
\eea
If we give a proper $a=a(t)$ and therefore $H=H(t)$, Eq.~(\ref{YM13}) can be
regarded as a third
order differential equation with respect to $\lambda$. If we find the
solution of $\lambda$
with three constants of the integration, we find the explicit form of
$P(\phi)=P(t)$ by using Eq.~(\ref{YM12}) and further obtain $Q(\phi)$ by 
using Eq.~(\ref{YM9}). 
Thus, we find the explicit form of three parameter families of 
the action (\ref{YM2}). 
This tells that almost arbitrary time development of the university could 
be realized by the action (\ref{YM2}) or (\ref{YM1}). 

As an example, we may consider the case of the power law expansion: 
$
a=\left(t/t_1\right)^{h_1} 
$, 
$
H=h_1/t
$, 
where $t_1$ and $h_1$ are constants. By assuming 
$
\lambda = \left(h_1 - 1\right) \ln \left( t/t_1 \right) +
\lambda_1
$, 
where $\lambda_1$ is a constant, Eq.~(\ref{YM3}) is reduced to the algebraic 
equation: 
$
0=\left[2h_1/\left(h_1-1\right)\right] \bar{X}^2 + 
\left( -4h_1^2 + 13 h_1 + 2 \right) \bar{X}
+ \left(h_1 -1 \right)^2 \left(h_1 - 2\right)\left(4H_1 -20\right)
$, 
where $\bar{X}=\bar{\alpha}^2 t_1^2 \e^{2\lambda}$. 
If this equation has a real positive solution with respect to $\bar{X}$, 
we obtain $\lambda_1$ and therefore the exact form of $\lambda$. 
Consequently, we can reconstruct a model to give the above power 
expansion. 
Similarly, any other evolutional history of the universe may be reproduced 
by the specific form of the action under consideration.

\section{Finite-time future singularities in non-minimal Maxwell-Einstein 
theory}

In this section, we consider non-minimal Maxwell-Einstein theory with 
a general gravitational coupling. 
The model action is as follows~\cite{Bamba:2008ut}:
\begin{eqnarray}
S_{\mathrm{GR}} \Eqn{=}
\int d^{4}x \sqrt{-g}
\left( 
{\mathcal{L}}_{\mathrm{EH}}
+{\mathcal{L}}_{\mathrm{EM}} \right)\,,
\label{eq:6.1} \\ 
{\mathcal{L}}_{\mathrm{EM}}
\Eqn{=}
 -\frac{1}{4} I(R)
F_{\mu\nu}F^{\mu\nu}\,,
\label{eq:6.3}
\end{eqnarray}
where 
${\mathcal{L}}_{\mathrm{EH}}$ is the Einstein-Hilbert action in 
(\ref{eq:2.2}), 
$F_{\mu\nu} = {\partial}_{\mu}A_{\nu} - {\partial}_{\nu}A_{\mu}$
is the electromagnetic field-strength tensor, $A_{\mu}$ is the $U(1)$ 
gauge field, and $\tilde{I}(R)$ is an arbitrary function of $R$. 

Taking the variations of the action Eq.\ (\ref{eq:6.1}) with respect to the 
metric $g_{\mu\nu}$ and the $U(1)$ gauge field $A_{\mu}$, we obtain the 
gravitational field equation and the equation of motion of $A_{\mu}$ 
as~\cite{Bamba:2008ja}
\begin{eqnarray}
R_{\mu \nu} - \frac{1}{2}g_{\mu \nu}R
\Eqn{=} \kappa^2 T^{(\mathrm{EM})}_{\mu \nu}\,,
\label{eq:6.5} \\
T^{(\mathrm{EM})}_{\mu \nu}
\Eqn{=} 
I(R) \left( g^{\alpha\beta} F_{\mu\beta} F_{\nu\alpha}
 -\frac{1}{4} g_{\mu\nu} F_{\alpha\beta}F^{\alpha\beta} \right)
\nonumber \\
&&
{}+\frac{1}{2} \biggl[ I^{\prime}(R)
F_{\alpha\beta}F^{\alpha\beta} R_{\mu \nu}
+ g_{\mu \nu} \Box \left( I^{\prime}(R)
F_{\alpha\beta}F^{\alpha\beta} \right)
 - {\nabla}_{\mu} {\nabla}_{\nu}
\left( I^{\prime}(R)
F_{\alpha\beta}F^{\alpha\beta} \right)
\biggr]\,,
\label{eq:6.6} \\
&&
\hspace{-45.5mm}
\mathrm{and} 
\hspace{15mm}
 -\frac{1}{\sqrt{-g}}{\partial}_{\mu}
\left( \sqrt{-g} I(R) F^{\mu\nu}
\right) = 0\,,
\label{eq:6.7}
\end{eqnarray}
respectively,
where
$T^{(\mathrm{EM})}_{\mu \nu}$ is the contribution to
the energy-momentum tensor from the electromagnetic field.

It follows from Eq.~(\ref{eq:6.7}) that in the flat FRW background, 
the equation of motion for the $U(1)$ gauge field 
in the Coulomb gauge, ${\partial}^jA_j(t,\Vec{x}) =0$, and the case of 
$A_{0}(t,\Vec{x}) = 0$, is the same form as in Sec.~II. 
We therefore obtain the approximate solution as 
$\left|A_i(k,\eta)\right|^2 = |C(k)|^2$, where 
$|C(k)|^2$ is given by Eq.~(\ref{eq:A-4}). 
Using this solution, we find that the amplitude of the proper magnetic fields 
in the position space is given by 
$
|{B_i}^{(\mathrm{proper})}(t)|^2 =
\left(k|C(k)|^2/\pi^2\right) \left(k^4/a^4\right)
$ 
on a comoving scale $L=2\pi/k$. From this equation, we see that the proper 
magnetic fields evolve 
as $|{B_i}^{(\mathrm{proper})}(t)|^2 = |\bar{B}|^2/a^4$, where
$|\bar{B}|$ is a constant. 
The conductivity of the universe ${\sigma}_\mathrm{c}$
is negligibly small during inflation, because there are few charged particles
at that time. After the reheating stage, a number of charged particles are
produced, so that the conductivity immediately jumps to a large
value:~${\sigma}_\mathrm{c} \gg H$.
Consequently, for a large enough conductivity at the reheating stage,
the proper magnetic fields behave in proportion to $a^{-2}(t)$ in the
radiation-dominated stage and the subsequent matter-dominated
stage~\cite{Ratra:1991bn}. 

In this case, it follows from Eq.~(\ref{eq:6.6}) that 
the quantities corresponding to the effective energy density of the universe
$\rho_\mathrm{eff}$ and the effective pressure 
$p_\mathrm{eff}$ are given by
\begin{eqnarray}
\hspace{-15mm}
&&
\rho_\mathrm{eff} =
\left\{ \frac{I(R)}{2} + 3\left[
 -\left( 5 H^2 + \dot{H} \right) I^{\prime}(R) +
6 H \left( 4H\dot{H} + \ddot{H} \right) I^{\prime\prime}(R)
\right]
\right\}
\frac{|\bar{B}|^2}{a^4}\,,
\label{eq:6.13} \\
\hspace{-15mm}
&&
p_\mathrm{eff} = \biggl[ -\frac{I(R)}{6} - 
\left(H^2 - 5\dot{H} \right) I^{\prime}(R) + 
6 \left(20H^2\dot{H} - 4\dot{H}^2 + H\ddot{H} - \dddot{H} \right)
I^{\prime\prime}(R) 
-36\left( 4H\dot{H} + \ddot{H} \right)^2 I^{\prime\prime\prime}(R)
\biggr]
\frac{|\bar{B}|^2}{a^4}\,. 
\label{eq:6.14}
\end{eqnarray} 

We now suppose that $I(R)$ is (almost) constant at the present time, 
and that for the small curvature, $I(R)$ behaves as 
$
I(R) \sim I_0 R^\alpha 
$, 
where $I_0$ and $\alpha$ are constants. We consider the case $\alpha < 0$. 
The energy density of the magnetic fields is given by
$\rho_B = \left(1/2\right) |{B_i}^{(\mathrm{proper})}(t)|^2 I(R) = 
\left[|\bar{B}|^2/\left(2a^4\right)\right]I(R)$. 
We take de Sitter background as the future universe.
In such a case, when $R$ becomes much smaller in the future, 
the energy density of the magnetic field becomes larger and larger 
in comparison with its current value. 
Thus, the strength of current magnetic fields of 
the universe may evolve to very large values in the future universe.

In the flat FRW background, the Einstein equations are given by 
$
3H^2/\kappa^2 = \rho_\mathrm{eff}
$
and 
$
-\left(2\dot H + 3H^2\right)/\kappa^2 = p_\mathrm{eff}
$, 
where $\rho_\mathrm{eff}$ and $p_\mathrm{eff}$ are given by 
Eqs.~(\ref{eq:6.13}) and (\ref{eq:6.14}), respectively. 

We examine the form of $I(R)$ producing the Big Rip singularity 
$
H \sim h_0/\left(t_0 - t\right)
$, 
where $h_0$ is a positive constant, and $H$ diverges at $t = t_0$. 
In this case, the scalar curvature and the scale factor are given by 
$
R \sim 6h_0\left(2h_0 + 1\right)/\left(t_0 - t\right)^2 
$ and 
$
a \sim a_0 \left(t_0 - t\right)^{- h_0}
$, respectively, where $a_0$ is a constant.
We now assume that for the large curvature, $I(R)$ behaves as 
$I(R) \sim I_0 R^\alpha$. 
Hence $\rho_{\rm eff}$ in Eq.~(\ref{eq:6.13}) behaves as
$\left(t_0 - t\right)^{-2\alpha + 4h_0}$, but the left-hand side (l.h.s.) on 
the Friedmann equation $3H^2/\kappa^2 = \rho_\mathrm{eff}$ evolves as 
$\left(t_0 - t\right)^{-2}$. 
The consistency gives 
$- 2 = -2 \alpha + 4h_0$, i.e., 
$
h_0 = \left(\alpha -1\right)/2
$
or
$
\alpha= 1 + 2h_0
$. 
The Friedmann equation $3H^2/\kappa^2 = \rho_\mathrm{eff}$ also shows 
$3h_0^2/\kappa^2 = 
- \left[I_0 h_0 \left( 12h_0^2 + 6h_0 \right)^\alpha 
|\bar{B}|^2 \right] /\left(2 a_0^4\right)
$, 
where we have used $\alpha= 1 + 2h_0$. This equation 
requires that $I_0$ should be negative. As a result, 
it follows from $I(R) \sim I_0 R^\alpha$ and $\alpha= 1 + 2h_0$ that 
the Big Rip singularity can appear only when for the 
large curvature, $I(R)$ behaves as $R^{1+2h_0}$. If the form of $I(R)$ is 
given by other terms, the Big Rip singularity cannot emerge. 
We note that if exactly $I(R) = I_0 R^\alpha$,
$H=h_0/\left(t_0 - t\right)$ is an exact solution.

Next, we investigate the form of $I(R)$ giving a more general singularity 
$
H \sim h_0 \left(t_0 - t\right)^{-\beta}
$. 
In this case, the scalar curvature and the scale factor are given by 
$
R \sim 6h_0 \left[\beta + 2h_0 \left( t_0 -t \right)^{-\left(\beta-1 \right)}
\right] \left( t_0 -t \right)^{-\left(\beta+1 \right)}
$ 
and 
$
a \sim a_0 \exp \left[ h_0 \left(\beta-1\right)^{-1} 
\left( t_0 -t \right)^{-\left(\beta -1 \right)} \right] 
$, respectively. 

In Ref.~\cite{Nojiri:2005sx}, the finite-time future 
singularities has been classified in the following way:
\begin{itemize}
\item Type I (``Big Rip'') : For $t \to t_s$, $a \to \infty$,
$\rho \to \infty$ and $|p| \to \infty$. This also includes the case of
$\rho$, $p$ being finite at $t_s$.
\item Type II (``sudden'')~\cite{sudden} : For $t \to t_s$, $a \to a_s$,
$\rho \to \rho_s$ and $|p| \to \infty$
\item Type III : For $t \to t_s$, $a \to a_s$,
$\rho \to \infty$ and $|p| \to \infty$
\item Type IV : For $t \to t_s$, $a \to a_s$,
$\rho \to 0$, $|p| \to 0$ and higher derivatives of $H$ diverge.
This also includes the case in which $p$ ($\rho$) or both of $p$ and $\rho$
tend to some finite values, while higher derivatives of $H$ diverge.
\end{itemize}
Here, $t_s$, $a_s (\neq 0)$ and $\rho_s$ are constants.
We now identify $t_s$ with $t_0$.
The Type I corresponds to $\beta>1$ or $\beta=1$ case, Type II to $-1<\beta<0$
case, Type III
to $0<\beta<1$ case, and Type IV to $\beta<-1$ but $\beta$ is not any
integer number. 
We note that if only higher derivatives of the Hubble rate 
diverge, then some combination of curvature invariants also diverges and it 
leads to singularity. 

We assume that for the large curvature, $I(R)$ behaves as 
$I(R) \sim I_0 R^\alpha$. 
If $\beta <-1$, in the limit $t \to t_0$, $R \to 0$. We therefore consider 
this case later. 
If $\beta >1$, $a \to \infty$ and 
$\rho_\mathrm{eff} \to 0$ and $p_\mathrm{eff} \to 0$
because $\rho_\mathrm{eff} \propto a^{-4}$ and
$p_\mathrm{eff} \propto a^{-4}$. 
On the other hand, $H \to \infty$. 
Hence the Einstein equations cannot be satisfied.

If $\alpha > 0$ and $0< \beta <1$, $\rho_{\rm eff}$ in Eq.~(\ref{eq:6.13}) 
evolves as $\left(t_0 - t\right)^{-\alpha \left( \beta + 1 \right)}$, but the 
l.h.s. on $3H^2/\kappa^2 = \rho_\mathrm{eff}$ behaves as 
$\left(t_0 - t\right)^{-2\beta}$. 
The consistency gives 
$- 2\beta = -\alpha \left( \beta + 1 \right)$, i.e., 
$
\beta = \alpha/\left(2-\alpha\right) 
$
or
$
\alpha = 2\beta/\left(\beta+1\right)
$. 
 From $3H^2/\kappa^2 = \rho_\mathrm{eff}$, we also find 
$3h_0^2/\kappa^2 = 
- \left[ I_0 \left( 6h_0 \beta \right)^\alpha \left( 1-\beta \right)
|\bar{B}|^2 \right]/\left[ 2 a_0^4 \left( \beta + 1 \right)\right]
$, 
where we have used $\alpha = 2\beta/\left(\beta+1\right)$ 
and on the l.h.s. we have taken only the leading term. 
This equation requires that $I_0$ should be negative. 
If $\alpha > 0$ and $0< \beta <1$, in the limit $t \to t_0$, 
$a \to a_0$, $R \to \infty$, 
$\rho_\mathrm{eff} \to \infty$ and 
$|p_\mathrm{eff}| \to \infty$. Hence the Type III singularity appears. 
If $\alpha > 0$ and $-1< \beta <0$, $\rho_\mathrm{eff} \to \infty$, 
but $H \to 0$. Thus $3H^2/\kappa^2 = \rho_\mathrm{eff}$ cannot be satisfied. 

If $\left( \beta-1 \right)/\left( \beta+1 \right) < \alpha <0$ and 
$-1< \beta <0$, in the limit $t \to t_0$,
$a \to a_0$, $R \to \infty$, $\rho_\mathrm{eff} \to 0$ and 
$|p_\mathrm{eff}| \to \infty$. 
Although the final value of $\rho_\mathrm{eff}$ is not finite but 
vanishes, this singularity can be considered to the Type II.
The reason is as follows. 
In this case, when $I$ and $H$ are given by 
$I = 1 + I_0 R^\alpha$ and $H=H_0+h_0 \left(t_0 -t\right)^{-\beta}$, 
respectively, where $H_0$ is a constant, 
in the above limit $\rho_\mathrm{eff} \to \rho_0$. From 
Eq.~(\ref{eq:6.13}) and $3H^2/\kappa^2 = \rho_\mathrm{eff}$, we find 
$\rho_0 = 3H_0^2/\kappa^2 = 
|\bar{B}|^2/\left(2 a_0^4\right)$. Hence $\rho_0$ is a finite value. 

If $\alpha \leq \left( \beta-1 \right)/\left( \beta+1 \right)$ and 
$-1< \beta <0$, in the limit $t \to t_0$, 
$a \to a_0$, $R \to \infty$, 
$\rho_\mathrm{eff} \to 0$ and $|p_\mathrm{eff}| \to 0$, 
but $\dot{H} \to \infty$. 
Hence 
$-\left(2\dot H + 3H^2\right)/\kappa^2 = p_\mathrm{eff}$ 
cannot be satisfied. 
If $\alpha < 0$ and $0< \beta <1$, $\rho_\mathrm{eff} \to 0$, but 
$H \to \infty$. Thus $3H^2/\kappa^2 = \rho_\mathrm{eff}$ cannot be satisfied. 

In addition, we investigate the case in which $\beta <-1$. In this case,
in the limit $t \to t_0$, $a \to a_0$ and $R \to 0$.
We assume that for the small curvature, $I(R)$ behaves as 
$I(R) \sim I_0 R^\alpha$. 
If $\alpha \geq \left( \beta-1 \right)/\left( \beta+1 \right)$,
in the limit $t \to t_0$, $\rho_\mathrm{eff} \to 0$,
$|p_\mathrm{eff}| \to 0$, and higher derivatives of $H$
diverge. Hence the Type IV singularity appears. 
If $0 < \alpha < \left( \beta-1 \right)/\left( \beta+1 \right)$,
$\rho_\mathrm{eff} \to 0$ and $|p_\mathrm{eff}| \to \infty$.
However, $H \to 0$ and $\dot{H} \to 0$. Thus 
$-\left(2\dot H + 3H^2\right)/\kappa^2 = p_\mathrm{eff}$ 
cannot be satisfied.

We remark that 
if $I(R)$ is a constant (the case in which $I(R)=1$ corresponds to the 
ordinary Maxwell theory), any singularity cannot appear. 
We also mention the case in which $I(R)$ is given by the Hu-Sawicki 
form in Eq.~(\ref{eq:2.49}) as $I(R)=f_{\mathrm{HS}}(R)$ 
or expressed by Eq.~(\ref{eq:3.10}) as $I(R)=f_{\mathrm{NO}}(R)$. 
If $\beta < -1$ and $I(R)=f_{\mathrm{HS}}(R)$ or $I(R)=f_{\mathrm{NO}}(R)$, 
in the limit $t \to t_0$, 
$a \to a_0$, $R \to 0$, $\rho_\mathrm{eff} \to 0$ and 
$|p_\mathrm{eff}| \to 0$. 
In addition, higher derivatives of $H$ diverge.
Thus the Type IV singularity appears. 

As a consequence, it has been demonstrated that the Maxwell theory coupled 
non-minimally with the Einstein gravity may produce finite-time singularities 
in future, depending on the form of the non-minimal gravitational coupling. 
The general conditions for $I(R)$ in order that the finite-time future
singularities cannot appear are that in the limit 
$t \to t_0$, $I(R) \to \bar{I}$, 
where $\bar{I} (\neq 0)$ is a finite constant, 
$I^{\prime}(R) \to 0$, $I^{\prime\prime}(R) \to 0$ and 
$I^{\prime\prime\prime}(R) \to 0$.

\section{Conclusion}

In the present article, we have reviewed 
cosmology in non-minimal YM/Maxwell theory, in which the YM 
(electromagnetic) field couples to a function of the scalar curvature. 
It has been shown that power-law inflation can be realized due to 
the non-minimal gravitational coupling of the YM field which may be caused 
by quantum corrections. 
Furthermore, we have considered non-minimal vector model in the framework of 
modified gravity. It has been demonstrated that both inflation and 
the late-time accelerated expansion of the universe can be realized. 
We have also discussed the cosmological reconstruction of the YM theory. 
In addition, we have studied late-time cosmology in the non-minimal 
Maxwell-Einstein theory. We have investigated the forms of the non-minimal 
gravitational coupling which generates the finite-time future singularities 
and the general conditions for this coupling in order that the finite-time 
future singularities cannot appear. 
Finally, it is interesting to mention that using the reconstruction method 
developed in Refs.~\cite{Nojiri:2006be, reconstruction method}, 
one can present models with the crossing of the phantom divide 
in non-minimal YM-modified gravity generalizing the models suggested in 
Ref.~\cite{Bamba:2008hq}. 

\section*{Acknowledgments} 
We deeply appreciate the invitation of Professor P.~M.~Lavrov and Professor 
V.~Ya.~Epp to submit this article to the anniversary volume 
\textit{The Problems of Modern Cosmology} 
on the occasion of the 50th birthday of Professor Sergei D. Odintsov. 
We also thank Professor S.~D.~Odintsov for his collaboration in 
Refs.~\cite{Bamba:2008xa, Bamba:2008ut} very much. 
In addition, we are grateful to Professor Misao Sasaki for very helpful 
discussion of related problems. 
The work by S.N. is supported in part by the Ministry of Education, 
Science, Sports and Culture of Japan under grant no.18549001 and Global
COE Program of Nagoya University provided by the Japan Society
for the Promotion of Science (G07). 
The work by K.B. is supported in part by 
National Tsing Hua University under Grant \#: 97N2309F1.


\end{document}